\documentclass[amssymb,amsmath,pra,twocolumn,floatfix]{revtex4}
\usepackage{color}
\usepackage{graphicx}
\usepackage{amssymb}
\usepackage{amsmath}
\usepackage{array}

\newcommand{\bra}[1]{{\left\langle{#1}\right\vert}}
\newcommand{\ket}[1]{{\left\vert{#1}\right\rangle}}

\begin{document}
\title{Fault-Tolerant High Level Quantum Circuits: Form, Compilation and Description}

\author{Alexandru Paler$^1$, Ilia Polian$^2$, Kae Nemoto$^3$ and Simon J. Devitt$^{4}$}
\affiliation{$^1$Johannes Kepler University Linz, Altenberger Str. 69, 4070 Linz, Austria
.}
\affiliation{$^2$University of Passau, Innstr. 43, 94032, Passau, Germany}
\affiliation{$^{3}$National Institute of Informatics, 2-1-2 Hitotsubashi, Chiyoda-ku, Tokyo, Japan.}
\affiliation{$^{4}$Center for Emergent Matter Science, RIKEN, Wakoshi, Saitama 315-0198, Japan.}
\date{\today}

\begin{abstract}
Fault-tolerant quantum error correction is a necessity for any quantum architecture destined to tackle interesting, 
large-scale problems. Its theoretical formalism has 
been well founded for nearly two decades.  However, we still 
do not have an appropriate compiler to produce a fault-tolerant, error corrected description from a higher level quantum 
circuit for state of the art hardware models.  There are 
many technical hurdles, including dynamic circuit constructions that occur when 
constructing fault-tolerant circuits with commonly used error correcting codes.  We introduce a package that 
converts high level quantum circuits consisting of commonly used gates 
into a form employing all decompositions and ancillary protocols needed for fault-tolerant error 
correction.  We call this 
form the (I)initialisation, (C)NOT, (M)measurement form (ICM) and consists of an initialisation layer of 
qubits into one of four distinct states, a massive, deterministic array of CNOT operations and a series of time 
ordered $X$- or $Z$-basis measurements. The form allows a more flexbile approach towards circuit optimisation. At the same time, the package outputs a standard circuit or a canonical geometric description which is a necessity for operating current state-of-the-art hardware architectures using topological quantum codes.
\end{abstract}
\maketitle

The construction and compilation of quantum algorithms has long been an active area of research 
\cite{K96+,O00,BCS03,S04,T04,AG05,SV06,GLRSV13,WS14,DMN08}.  Since the first work on 
designing circuits from abstract quantum algorithms \cite{BBCD95,NC00,KC00,Y93,VBE96,PMH08,VI05,SPK13}, 
research has been centred on many different aspects of compilation and optimisation.  These techniques have focused on aspects such as 
minimising the total number of qubits \cite{B03,SMB04,M07,OM12}, the depth of quantum circuits \cite{HS05,BK09,AMMR12,AWSD13}, 
constraints related to hardware \cite{FDH04,DKRS06,M07,SFH07,SWD10,SSP13,WKS15} and the minimisation of certain quantum 
gates that are resource intensive when implemented in  
error corrected hardware \cite{S12,GS12,RS14,KMM13,WBS14,GKMR14}.  
However, despite the significant quantity of research in this area, much of the early work omitted details
 of fault-tolerant error correction protocols.  These protocols are not only instrumental in implementing 
any quantum algorithm, beyond a handful of qubits, but place constraints on the cost metrics used 
when assessing optimal circuit constructions for a higher level algorithm \cite{DSMN13}.  

In recent years, theoretical research has begun to take into account the considerations of error corrected computation. 
Both high level quantum programming languages and efforts in optimisation have attempted to re-engineer circuits under more 
appropriate constraints.  Most notably is the increased number of papers recognising that in fault-tolerant quantum computation 
a Clifford + $T$ universal gate set is generally used and the resource penalty associated with the single qubit $T = \ket{0}\bra{0} + e^{i\pi/4}\ket{1}\bra{1}$ gate requires 
the minimisation of both the number of such gates and the corresponding $T$-depth (the number of sequential applications of 
$T$-gates in a circuit) \cite{S12,GS12,RS14,KMM13,WBS14,GKMR14}.  
However, much of this work still abstracts out concrete choices 
of quantum error correction (QEC) coding and/or specific hardware implementations when compiling and optimising quantum circuits.  

From a hardware perspective, topological codes have emerged as the preferred method for achieving a scalable design.  The majority 
of scalable architectures proposed either use the 2-dimensional surface code \cite{DKLP02,FSG08,FMMC12}, 
or the 3-dimensional Raussendorf model \cite{RHG07,FG08} as the underlying error correction method.  These models are 
preferred as they can be defined on a 2- or 3-dimensional nearest neighbour array of qubits, facilitating experimental 
fabrication, and have one of the highest fault-tolerant thresholds of any error correction scheme (approaching 1\% 
under certain circuit assumptions) \cite{WFH11,S14}.  Additionally, realisation of error corrected algorithms is a function 
of how the array of qubits is measured, 
rather than how it is constructed (i.e. the actual choice of algorithm only dictates the physical size of the lattice needed at the 
hardware level, nothing more) \cite{DN12,DSMN13,FDJ13}, with error correction decoding and correction occurring concurrently with algorithmic implementation \cite{D10,W13}.  In recent years 
numerous results have been published demonstrating hardware accuracy sufficient for this coding model in both ion traps and 
superconducting systems \cite{DFSG08,JMFMKLY10,YJG10,NTDS13,LWFMDWH15}.

These topological codes also have the property, along with more traditional concatenated CSS codes, that a universal, error 
corrected gate set is built using the Clifford + $T$ library.  The inherent fault-tolerant operations allowed in these models are 
initialisations/measurements in the $Z$- and $X$-bases and a two qubit CNOT gate.  Hadamard ($H = (X+Z)/\sqrt{2}$) and Phase ($P = \sqrt{Z}$) 
gates can be realised in multiple ways \cite{BK05+,F12,A07,J13+},  
to complete the Clifford set, while the $T$-gate is realised through state injection, purification 
protocols and teleportations \cite{BK05+}.  State distillation protocols, necessary for the $T$ gate and some 
realisations of the $H$- and $P$-gates, dominate the resources required for a large scale computation \cite{DSMN13,FMMC12}.

\section{Results}

In this work we present a decomposition and compilation package that can be the \emph{starting point} for global 
optimisation of a fault-tolerant, error corrected circuit in either the topological models or any other model of error correction 
where a Clifford + $T$ library is used. This is enabled by bringing high level quantum circuits consisting of commonly used gates 
into the ICM form, which employs all decompositions and ancillary protocols needed for fault-tolerant error 
correction.

The compiler outputs a circuit description necessary for the execution of the high level circuit on a large scale quantum computer using topological QECs. The source code for this compiler can be found at \url{https://github.com/alexandrupaler/icmconvert}.\\

In the appendix we provide detailed explanations about a formalism for time-optimal computation that allows dynamic quantum circuit corrections to be 
converted into a deterministic circuit \cite{F12}, each stage of the compilation process, including standard circuit identities used \cite{NC00}, gate approximations based on the results from Ross and Selinger \cite{RS14}, and fault-tolerant ancillary protocols such as state injection and 
distillation \cite{BK05+}.

\section{Fault-tolerant quantum circuit form}

The form of fault-tolerant quantum circuits, which we call the 
(I)nitialisation, (C)not, (M)easurement form \cite{PPKD15}, is the result of converting a high level quantum circuit. A high level circuit description is the equivalent of an abstract high level quantum algorithm converted into a standard quantum circuit representation utilising a well defined universal gate set. The quantum algorithm, described using Quipper \cite{GLRSV13} or Liquid \cite{WS14}, is output into a circuit using those tools and languages, and compiled into an ICM high level fault-tolerant quantum circuit using the herein presented framework. The ICM form is presented by employing the formalism of time optimal quantum computation \cite{F12+} and recent advances in Clifford + $T$ gate decompositions \cite{RS14}.  

This structure of the form consists of three layers. The first is a series of qubit initialisations in either the $\ket{0}$ state, 
the $\ket{+}$ state or the ancillary states $\ket{A} = \ket{0}+e^{i\pi/4}\ket{1}$, $\ket{Y} = \ket{0} + i\ket{1}$. The $\ket{A}$ and $\ket{Y}$ states are 
used to implement teleported $P = \sqrt{Z}$ and $T = \sqrt{P}$ gates.  The second layer is a massive array of purely CNOT gates which 
implement both the high level decomposed algorithm and \emph{all} ancillary protocols for fault-tolerant computation.  
The final layer is a \emph{time-ordered} series of $X$- and $Z$ basis measurements.

Unlike standard measurement based computation, the ICM representation only chooses between $X$- and $Z$-basis measurements.  This is an important distinction, because non-Clifford basis measurements require an extensive ancillary overhead due to fault-tolerant error correction requirements.  

ICM allows for a deterministic array of qubit initialisations and CNOT gates to describe {\em any} 
high level quantum algorithm, incorporating all necessary fault-tolerant protocols. The standard implementation of Clifford 
+ $T$ circuits involves active $P$-gate corrections that must be adaptively patched into the quantum circuit as it is executed.  
A correction is required 50\% of the time for each $T$ gate.

Optimising circuits beyond individual $T$ gates is 
problematic as we either need to reserve resources for {\em possible} corrections or attempt to optimise the 
exponential number of possibilities for circuits containing many $T$-gates.  
The time-optimal approach to Clifford + $T$ circuits folds this indeterminism into whether certain qubits 
are measured in the $X$- or $Z$-basis. The approach provides a deterministic initialisation and CNOT structure for an arbitrary large 
circuit. 

Hence from this formalism, it is now possible to optimise circuits starting from any point, from individual $T$-gates with 
required ancillary protocols up to the entire high level circuit.\\


\section{Compilation}

\begin{figure}
	\includegraphics[width=\columnwidth]{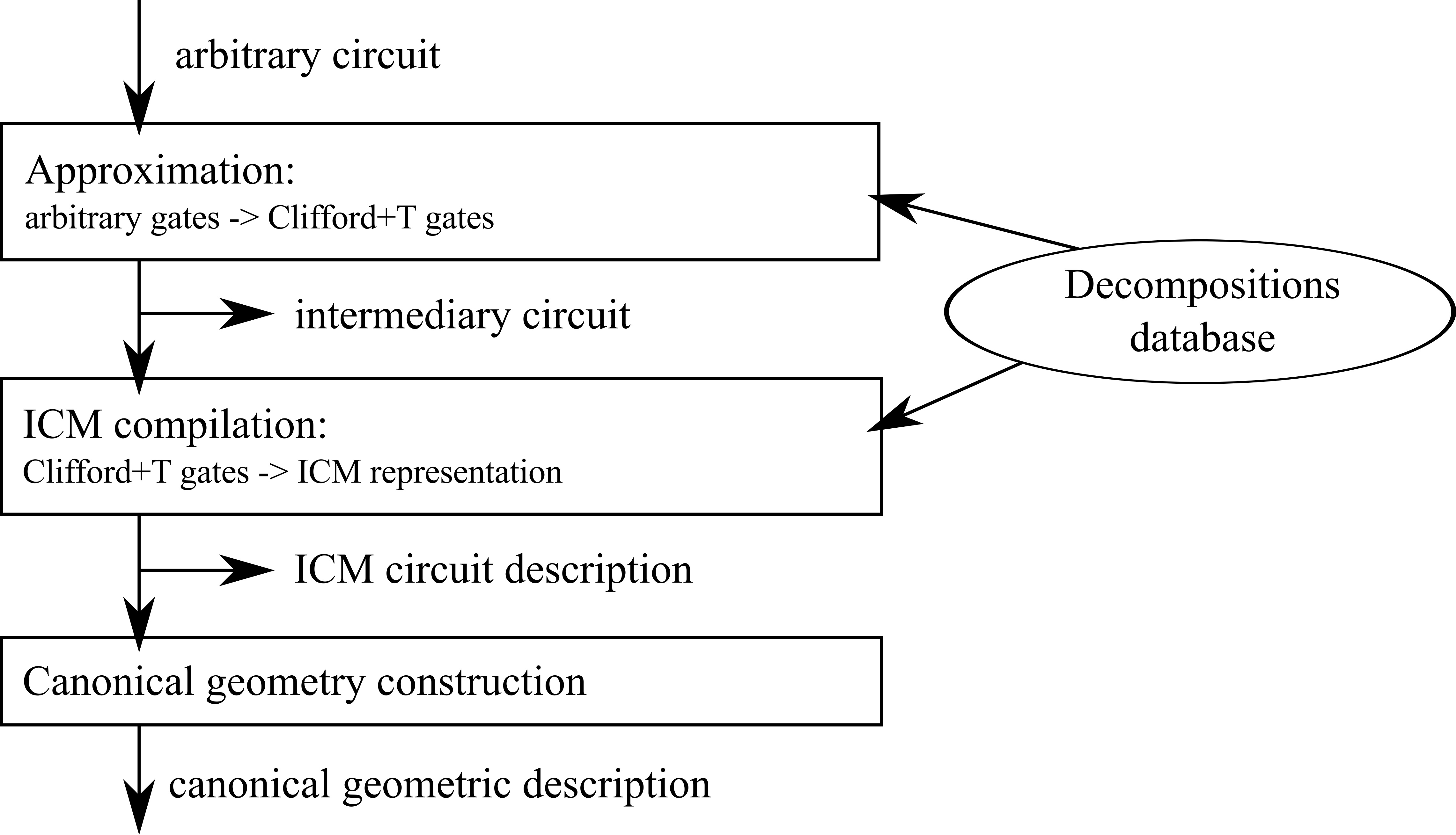}
	\caption{The ICM compiler work flow.}
	\label{fig:arch}
\end{figure}

The ICM form of arbitrary high level quantum circuits is obtained by using an ICM compiler. This takes an arbitrary quantum circuit as input. It outputs a standard circuit description and a canonical geometric structure, also called geometric description, that represents its implementation for topological QEC codes. Figure \ref{fig:arch} illustrates the compiler work flow. The following will detail the steps utilised in the compiler to allow for a fault-tolerant compatible description (quantum circuit or geometric) of the higher level circuit.\\

It should be noted that this compiler is acting at the {\em encoded} level, i.e. each qubit in the circuit description is encoded with an 
appropriate QEC. The physical details of how individual qubits are manipulated are assumed to 
be compatible with the QEC implementation (i.e. physical gate can be switched on or off and have error rates low enough to satisfy the fault-tolerant 
threshold of the QEC code).

The {\em arbitrary circuit} input to this workflow is a quantum algorithm that is already decomposed into a standard circuit gate set consisting of gates such as Toffoli, arbitrary single qubit rotations and CNOT gates. This circuit may be optimised for qubits and/or computational time before further decompositions.  This work does not address this level of circuit level optimisation.  Once a circuit has been specified (under whatever constraints), it is input into this workflow which will apply further decompositions to ultimately reach a fault-tolerant, error-corrected compatible representation.

The input circuit is specified as a list of non-ICM gates (the arbitrary circuit in Figure \ref{fig:arch}) and the ICM form is algorithmically constructed by replacing all non-ICM gates with their corresponding fault-tolerant ICM implementations. Arbitrary non-ICM circuits include gates from a previously chosen universal gate set. There are two possible situations: either the universal gate set is finite (e.g. Clifford + $T$ decompositions of a Toffoli gate) or infinite (Arbitrary single qubit rotations are {\em approximated} up to a 
residual error rate).

As a consequence, compiling an ICM circuit defined over an infinite gate set is a two-step process. Firstly, our software uses in the \emph {approximation step} (first layer in Figure~\ref{fig:arch}) as detailed in Ref. \cite{RS14} and outputs an intermediate circuit composed of gates from the finite (approximate) universal gate set $Q1=\{H,P,T\}$. Secondly, during the \emph{compilation step}, the intermediate circuit is further decomposed into the ICM representation (second layer in Figure~\ref{fig:arch}). For example, after decomposing an arbitrary $Z$-rotational gate into a gate sequence from $Q1$, the $H$,$P$,$T \in Q1$ gates are compiled into ICM.

For the compilation step we chose to use the $\{H, T, T^\dagger, P, P^{\dagger}, CNOT\}$ finite gate set (Clifford + $T$, where $T^{\dagger}$ and 
$P^\dagger$ are realised by re-interpreting measurement results of an ICM circuit), but treat the decomposition of the Toffoli, CPHASE, Controlled-V, Controlled-V$^\dagger$ gates as belonging to the approximation step. Although these gates have exact decompositions, their optimality in terms of $T$ count is still researched (e.g. the Toffoli gate \cite{J12,J13}).  The controlled-$V$ 
gate, where $V = \sqrt{X}$ is a finite controlled rotation as it does not require approximating arbitrary rotations and is heavily utilised in 
reversible circuit construction.  Arbitrary controlled rotations can also be converted into ICM form by using the identities shown in the appendix and 
then performing an approximation to arbitrary single qubit rotations using the algorithm of Ross and Selinger \cite{RS14}.

Figure \ref{fig:arch} mentions a decomposition database which includes known gate decompositions. The database stores the corresponding circuit representing the matching gate to be decomposed. For example, the Toffoli gate decomposition is loaded from the database during the approximation step. The $T$ gate ICM decomposition is loaded during the compilation step. The current version of our compiler uses a text based human readable database file format.

The resulting ICM circuit description includes a qubit initialisation list, a CNOT gate list and a measurement list (see output of second layer in Figure~\ref{fig:arch}). In general, a circuit's qubit initialisations and measurements are configurable because practical circuits are designed to support multiple input-output transformations. Exceptions exist for qubits used in specific protocols (e.g. distillation, teleportation) or ancillary workbench qubits (e.g. in adders). For example, as shown in the code segment below, in the circuit description of a $T$ gate (see the figure for the teleported $T$ gate in the appendix), the first line specifies the initialisation basis of the second qubit, the second line specifies a CNOT controlled by the second qubit and targeting the first qubit. The third line indicates that the first qubit is measured in the $Z$ basis. The initialisation of the first qubit is not specified, because it is an input and thus configurable. At the same time, the measurement of the second qubit is not specified.  Note:  For a circuit that represents an entire algorithm, there are no input or output qubits.  Every qubit is initialised and eventually measured.  Certain techniques in topological optimisation, specifically bridging \cite{FD12}, has significant power when we do not have to consider inputs/outputs which restrict which qubits can be bridged and which cannot.

{\scriptsize
\begin{verbatim}
init 2 A
cnot 2 1
measure 1 Z
\end{verbatim}
}

For input circuits using finite sets the complexity of the ICM transformation algorithm is linear in the number of gates in the input circuit. However, because gates from an infinite gate set have to be firstly approximated, the complexity of this procedure is a function of the chosen approximation method. In the specific case of this software, the number of gates needed scales logarithmically with the approximation error \cite{RS14}. The overhead associated with the ICM representation depends on the ICM compilation step, because the optimality of approximating arbitrary quantum gates is a function of the resulting $T$ gate count, which in turn is algorithm specific. For example, Ross and Selinger's algorithm achieves a $T$ count bounded by $4\log_2(1/\varepsilon) + \mathcal{O}(\log(\log(1/\varepsilon)))$ for an approximation precision $\varepsilon$ \cite{RS14}.

For a $q$-qubit non-ICM circuit consisting of $n$ gates chosen from the finite universal gate set, the resulting ICM circuit description (not including distillation protocols) will require $\mathcal{O}(q)$ qubits and $\mathcal{O}(n)$ gates, because the fault-tolerant teleportation-based gate constructions introduce a constant number of ancillary qubits (maximum 5 for the $T$ gate) and of gates (maximum 6 for the $T$ gate). Including distillation protocols into the ICM description introduces a polynomially bounded overhead of additional resources (qubits and gates).\\
\\

\section{Geometric description}
Topologically error-corrected computations have a visual representation. The canonical geometric description is a three-dimensional bijective mapping of the circuit description. Bijectivity is the property of each circuit description element to have a single corresponding structure in the geometric description, and vice versa. The geometric description can be optimised \cite{FD12,FDJ13,J13+,PF13}, but its canonicity and bijectivity would be difficult to regain. Similarly to the circuit description, the geometric description consists of three geometric regions: initialisation, CNOT and measurement. The bijective relation is recognised also in terms of the temporal axis: due to the fact that the geometry occupies a cuboid region of the three-dimensional space, the inputs are on the left of the cuboid, the outputs on the right, and CNOTs are ordered inside from left to right. This is exactly the quantum circuit formalism convention.

\begin{figure}
	\includegraphics[width=\columnwidth]{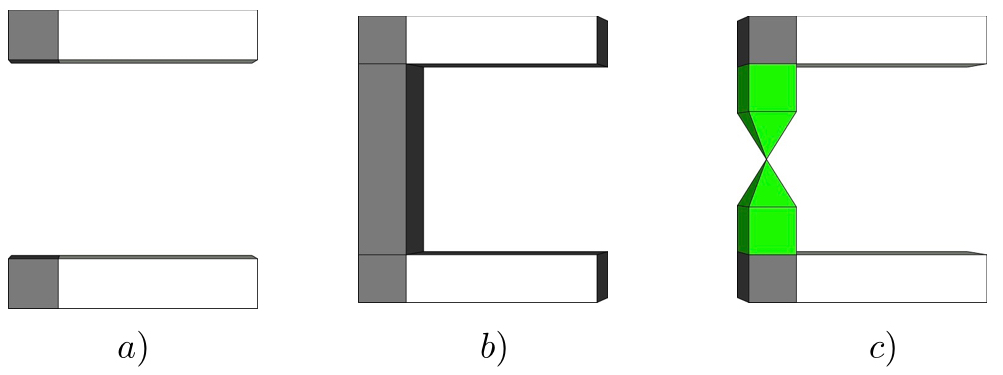}
\caption{{\bf Initialisation geometries for a primal qubit (two white strands): a)} $X$-basis initialisation; {\bf b)} $Z$-basis initialisation; {\bf c)} initialisation into $\ket{A}$ or $\ket{Y}$, depending on the state of the injected qubit whose existence is abstracted at the tip of the two pyramids.}
\label{fig:im}
\end{figure}

From a geometrical point of view, qubits are defined as pairs of strands (marked white in Figure \ref{fig:im}). There are two possible types of strands: primal and dual. These two types are a function of how the surface and Raussendorf codes work and are necessary to achieve topological braided logic \cite{RHG07,FMMC12}.  Consequently, two types of qubits can be defined: primal and dual. For example, a primal qubit is determined by a pair of primal strands. Due to the manner how strands are constructed in the surface code, they are also commonly called defects \cite{FMMC12}.  The difference between primal and dual defects corresponds to regions of the physical 
quantum hardware that is "switched off" in order to create them.  Restricting the discussion to the surface code \cite{FMMC12}, a primal 
defect is created when plaquette stabilisers are switched off and dual defects when vertex stabilisers are switched off.  

\begin{figure}
	\includegraphics[width=\columnwidth]{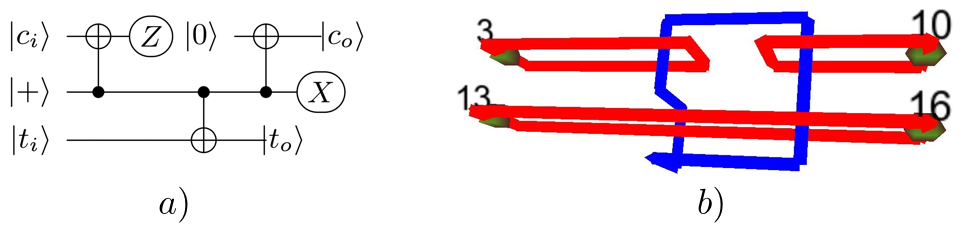}
\caption{{\bf A primal-primal CNOT:} {\bf a)} The circuit identity; {\bf b)} Primal strands are red and the duals are blue. The control qubit is represented in the upper region of the geometry, and the target qubit in the lower. By comparing the geometry with Figure \ref{fig:geomfig}a), the dual strand is identified as the ancillary initialised into $\ket{+}$ and measured in the $X$ basis. Each of the three braids between the blue dual and the red primal strands has a corresponding CNOT.}
\label{fig:geomfig}
\end{figure}

The CNOT gate is implemented, both in the surface code (2D) and Raussendorf code (3D), by braiding a primal strand (corresponding to the target qubit) around a dual strand (corresponding to the control qubit). This represents a primal-dual CNOT gate implementation, and can be extended to a primal-primal CNOT by using the circuit identity from Figure \ref{fig:geomfig}a). The qubits labelled with $\ket{c_i}$, $\ket{c_o}$, $\ket{t_i}$ and $\ket{t_o}$ will be mapped to primals, and the ancillary qubit (initialised into $\ket{+}$ and measured in the $Z$ basis) is going to be dual in the geometric description. The ancillary is the control for the three CNOTs, while the other qubits are targets. The resulting geometric representation of the dual qubit (Figure \ref{fig:geomfig}b)) hints at the vertical representation of the CNOT gate in the circuit formalism.

Strands are composed of multiple segments parallel to either the $x$, $y$ or $z$ axis \cite{RHG07,FG08}. Three-dimensional coordinates are sufficient to describe computations protected using the 2- or 3-dimensional codes. The properties of the coordinates are dictated by the structure of the surface codes: primal segment end point coordinates consist entirely of odd integer values, and dual segment end points of even integer values.

The generated circuit geometric description references qubit initialisations, measurements and the structure of the strands (list of segments). The braids do not have to be explicitly described, because, using geometric algorithms, it is possible to efficiently infer the braids (CNOTs) existing in a geometric description consisting of segments specified using their end point coordinates \cite{paler2014mapping}.

Geometrically described qubit initialisations are illustrated in Figure \ref{fig:im}. The measurement geometries are similar, but mirrored along the time axis (the x-axis for each figure). The interpretation of the geometries is interchanged between qubit types. For example, initialising a primal qubit into the $X$ basis uses the same geometry as dual qubit $Z$ initialisation. The primal $X$ initialisation (Figure \ref{fig:im}a) requires, from a geometrical perspective, to keep the strands separated, whereas a $Z$ initialisation requires connecting the strands into an U-shaped structure (Figure \ref{fig:im}b). Measuring a primal qubit into the previous two bases uses the same geometric structures, but vertically flipped due to the time axis assumed in the canonical representation. ICM circuit qubits can also be initialised into the $\ket{A}$ or $\ket{Y}$ states. The initialisation geometries are similar to primal $Z$-basis initialisation. The major difference is that instead of a single connecting strand, there is a point (injection point) from which two disjoint strands are defined (Figure \ref{fig:im}c). Between a qubit's strand end points there could be either no connecting strand, a connecting strand, or two disjoint strands intersecting at a point. The previous three options are valid for qubit initialisation and for measurements only the first options are valid in the context of this work.

\begin{figure}
\includegraphics[width=\columnwidth]{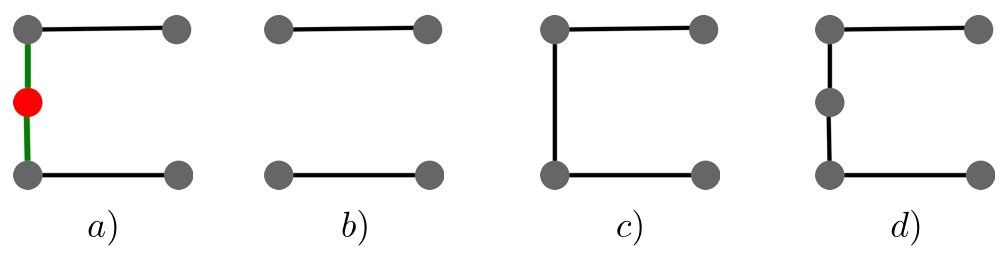}
\caption{The qubits represented by a canonical geometric description can be configured, similar to a circuit description, to be initialised or measured into multiple bases. The procedure requires a configuration point (red) and two segments (green) that connect to the end points (grey) of the qubit segments (black).}
\label{fig:l}
\end{figure}

Similar to the quantum circuit description, the geometric description is configurable at circuit inputs and outputs, too. We achieved this by introducing a configuration point at the geometric middle of an imaginary segment connecting the qubit's strands end points (Figure \ref{fig:l}a). The point corresponds to a possible injection point (compare Figures \ref{fig:l}a and \ref{fig:im}c). Configuring an ICM circuit to perform an $X$ measurement on a specific primal qubit is equivalent to deleting the corresponding point and the incident segments (Figure \ref{fig:l}b), while $Z$ measurement requires joining the incident segments and deleting the point (Figure \ref{fig:l}c). The same scenarios can be applied for $X$ or $Z$ initialisation, but the point and its incident segments are kept into the geometry for ancillary state initialisation (Figure \ref{fig:l}d).

The generated geometric description includes a list of three-dimensional end point coordinates and a list of segments existing between the end points. Furthermore, the initialisation and measurement points are marked accordingly: $\ket{A}$ initialisation points, $\ket{Y}$ initialisation points and ICM circuit inputs and outputs (configurable geometry). Table \ref{tbl:geomdesc} contains the description, split into five columns, of the geometry from Figure \ref{fig:geomfig}. The first value (4) indicates the number of configuration points. The following two lines indicate the number of three-dimensional coordinates (23) and segments (23). The coordinates, enumerated in the second and third columns, are of the form \texttt{id, x,y,z}, where \texttt{id} is a running index. In the first column, the line \texttt{3,10,13,16} lists the coordinate indices representing the four initialisation and measurement points. 

The remaining entries in the first column and the entire second column are index pairs indicating end points of geometric segments. The coordinates of primal configuration points contain an odd integer value, due to their definition at the middle of primal segments. For example, point \texttt{3} is the middle of segment \texttt{1,2}, the sum of segments \texttt{1,3} and \texttt{2,3}.

Finally, the fourth column specifies the configuration point types. The third (control) and tenth (target) points are inputs, while the thirteenth (control) and sixteenth (target) are corresponding to outputs (compare with Figures \ref{fig:geomfig}a and \ref{fig:geomfig}b).\\
\\

\begin{table}
\begin{tabular}{>{\ttfamily}l |>{\ttfamily}l|>{\ttfamily}l|>{\ttfamily}l|>{\ttfamily}l}
4 & 11,13 & 1,0,0,0 & 14,2,12,0 & 3,i\\\cline{1-1}
23 & 12,13 & 2,0,0,2 & 15,2,12,2 & 10,i\\
23 & 14,16 & 3,0,0,1 & 16,2,12,1 & 13,o\\\cline{1-1}
3,10,13,16 & 15,16 & 4,0,6,0 & 17,-1,9,1 & 16,o\\\cline{1-1}
1,3 & 11,14 & 5,0,6,2 & 18,-1,5,1&\\
2,3 & 12,15 & 6,0,8,0 & 19,1,5,1&\\
4,5 & 17,18 & 7,0,8,2 & 20,1,5,-1&\\
1,4 & 18,19 & 8,0,12,0 & 21,3,5,-1&\\
2,5 & 19,2 & 9,0,12,2 & 22,3,5,1&\\
6,7 & 20,21 & 10,0,12,1 & 23,3,9,1&\\
8,1 & 21,22 & 11,2,0,0&&\\
9,1 & 22,23 & 12,2,0,2&&\\
6,8 & 17,23 & 13,2,0,1&&\\
7,9&&&&
\end{tabular}
\caption{The geometric description of the primal-primal CNOT from Figure \ref{fig:geomfig}. Due to its length, the description is split into five columns.}
\label{tbl:geomdesc}
\end{table}

\begin{figure*}[ht!]
	\includegraphics[width=\textwidth]{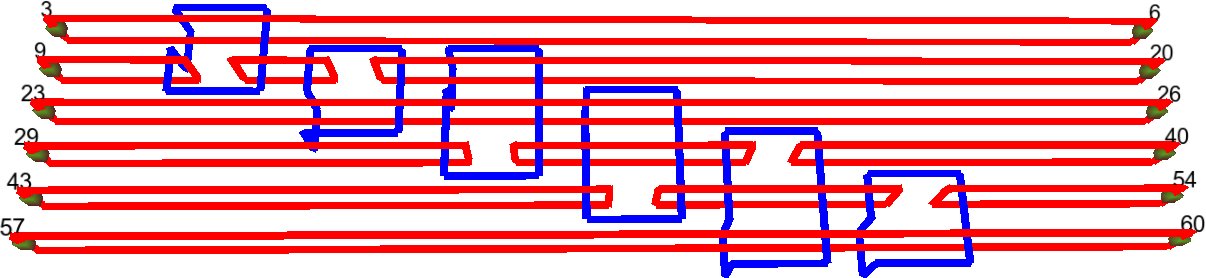}
	\caption{Example: the canonical geometry of the $T$ gate circuit (see appendix)}
	\label{fig:geomt}
\end{figure*}

\section{Example ICM circuits.}
Diagrammatically, even comparatively small quantum circuits (such as a single Toffoli gate) are extremely large, especially when 
state distillation protocols for $\ket{A}$ and $\ket{Y}$ states are introduced. The previously presented compiler can operate on arbitrary sized circuits and outputs to multiple formats (e.g. the geometric description).  

Due to the size restrictions, we illustrate two specific examples: a $T$ and a controlled-$V$ gate. Figure \ref{fig:geomt} illustrates the canonical geometric form for a topological circuit derived from the deterministic $T$ gate described in appendix, which is 
shown in ICM form).  In this circuit, input 3 and output 60 represent the actual input/outputs for the circuit while the remaining numbered nodes correspond to 
ancilla initialisations ($\ket{0}$, $\ket{+}$, $\ket{A}$ and $\ket{Y}$) and possible $X$- and $Z$-basis measurements to 
enact the teleported $T$-gate and possible $P$-gate correction. 

\begin{figure*}[ht!]
	\includegraphics[width=\textwidth]{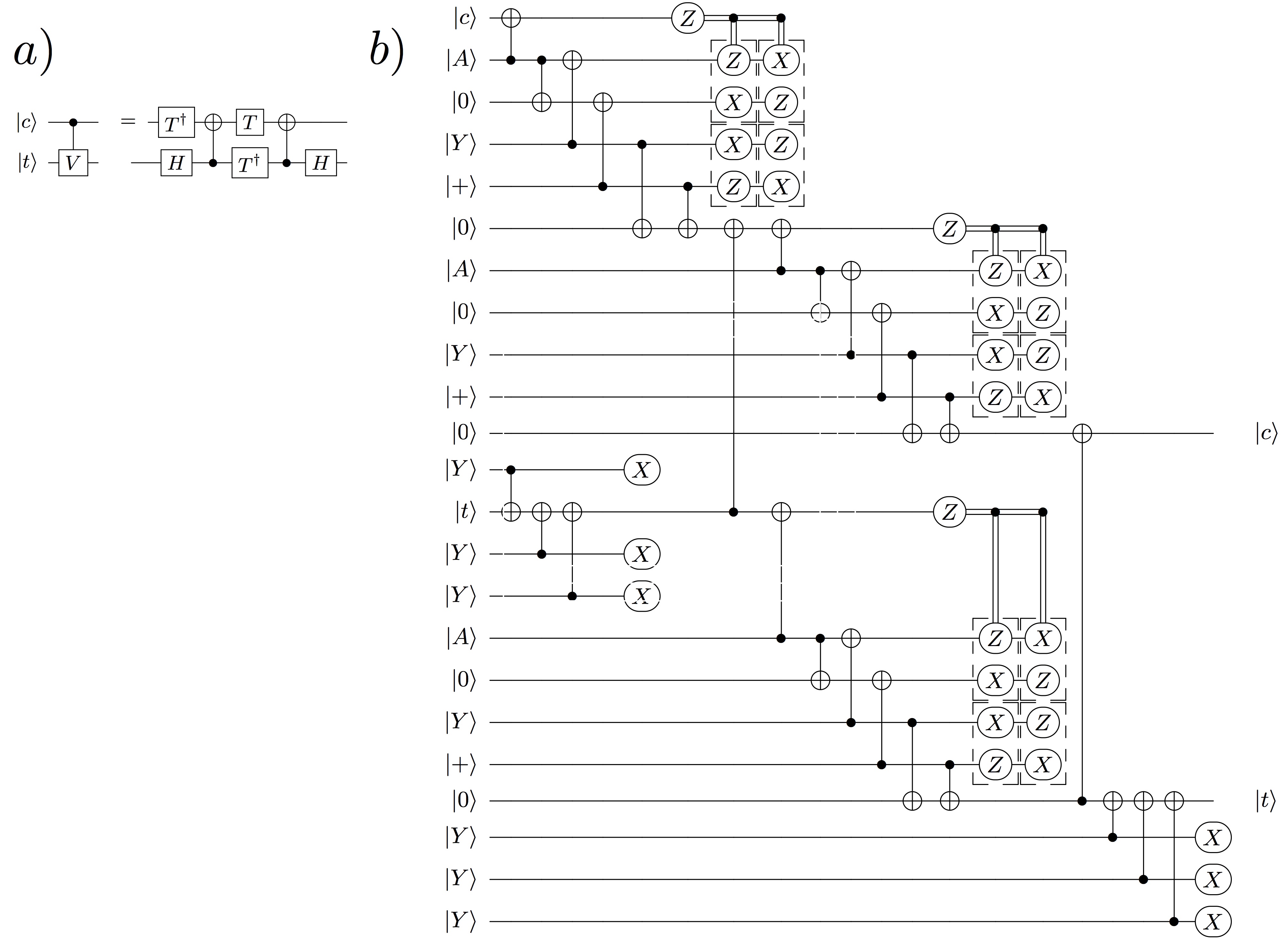}
\caption{{\bf Controlled-$V$ gate into ICM form. a)} Decomposition of a controlled-$V$ gate in terms of Clifford+ $T$ gates. {\bf b)} Set of decompositions for a controlled-$V$ gate, without the inclusion of state distillation circuits.  It is visible that, circuits expand significantly when placed into ICM form.  The circuit has a 
$T$-count of three and a $T$-depth of two, the two middle $T$-gates in the decomposition can be applied in parallel.}
\label{circ:Vgate1}
\end{figure*}

We also illustrate the series of decomposition steps to convert a 
controlled-$V$ gate over two qubits into ICM form {\em without including} 
distillation protocols. At the input level, a controlled-$V$ gate is a single two-qubit operation, as illustrate in Figure \ref{circ:Vgate1}a).  
This gate can be expressed exactly in the Clifford + $T$ form (which may or may not have already been performed before 
being compiled into ICM form) [Figure \ref{circ:Vgate1}b].  Compilation converts this circuit into the one shown in Figure \ref{circ:Vgate1}c), 
where $T$-gates are compiled into teleportation circuits with both $\ket{A}$ and $\ket{Y}$ ancillary states and each Hadamard gate 
is decomposed into three teleportations, each requiring a $\ket{Y}$ ancilla.  

The pair of dotted measurement boxes represents bases choices for qubit
measurement that are predicated on the logical measurement results from 
previous teleportations (the classical control dependance is indicated).  Temporal ordering is indicated by 
staggering each $T$-gate sub-circuits measurements.  This staggering is required in case classical $X$-corrections are required which 
can convert subsequent $T$-gates to $T^\dagger$-gates and vice versa.  This ensures that the $T$-depth of the ICM circuit is identical to 
the higher level Clifford + $T$ description.

\section{Discussion}

The ICM form allows for an arbitrary high-level circuit, obtained from high level quantum languages like e.g. Quipper, to be decomposed into a time-optimal, 
deterministic form.  Deterministic form implies that the array of CNOTs are fixed and the only probabilistic 
elements in the circuit are the $X/Z$ basis choices for measurements that occur through the decomposition of 
$T$-gates.  This form is, at least at the circuit description level, deliberately wasteful. We intentionally introduce a large number 
of ancillary qubits in order to have a deterministic array of qubit initialisations and CNOTs which can then be converted to 
a geometric structure (for topological codes) or to another output format if different error correction codes are used.  
This provides us with a starting point for further optimisation at any level.  Individual high-level gates (Toffoli gates 
or arbitrary logical rotations) could be optimised or the entire high-level quantum circuit, if desired.\\

\section{Optimal geometrical descriptions}

\begin{figure*}
	\includegraphics[width=2\columnwidth]{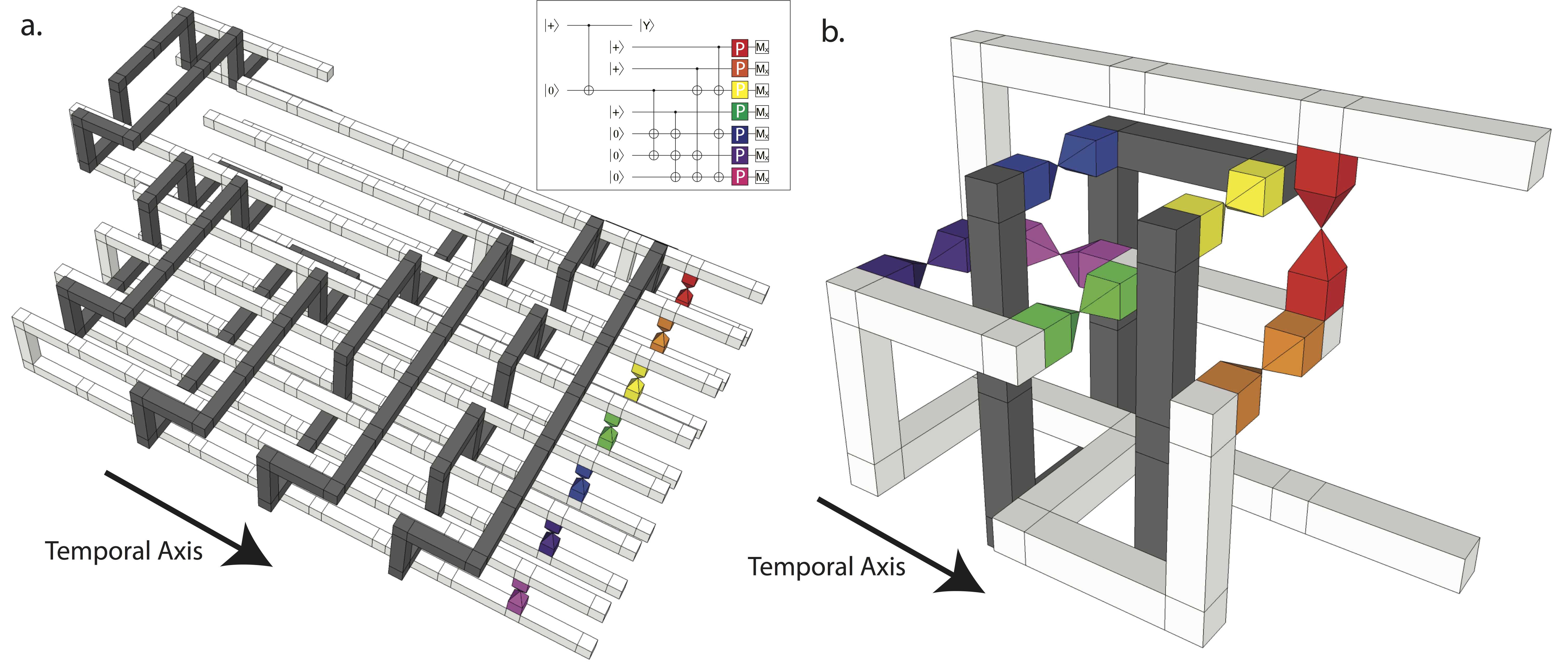}
	\caption{{\bf Example of resource optimisation for a topological quantum circuit \cite{FD12}.  a)} An ICM form of a state distillation circuit 
	is converted to a geometrical representation for topological codes.  {\bf b)} Compression techniques are then used to reduce 
	the space/time volume of the geometric structure which reduces physical resources for computation.  In this example, 
	the space/time volume was reduced by a factor of 11.}
	\label{fig:topological}
\end{figure*}

Restricting the discussion to topological quantum codes, namely the surface code and the Raussendorf code, it has been 
shown that various topological techniques can be used to significantly reduce the space/time volume of a quantum circuit \cite{FD12,FDJ13,J13+,PF13,PDF16}.  
Figure \ref{fig:topological} illustrates what has been derived previously for a distillation circuit for $\ket{Y}$ states.  
In this example, the space/time volume of a topological circuit was reduced by a factor of 11 from the original, 
canonical, ICM circuit.  This space/time optimisation relates directly to the number of 
physical qubits and computational time required to implement a circuit and is a purely classical problem \cite{DSMN13}.  
Compression of topological circuits result in structures that bare very little resemblance to the original circuit specification 
and consists of topological structures that no longer directly relate to individual qubits and CNOT gates in the 
original circuit \cite{FDJ13}.  Therefore, the fan-out of ancillary qubits in the ICM representation may not be detrimental to a 
compressed and optimised topological structure.  A formal solution to the compression of topological circuits has still 
not been developed and therefore the degree in which resources from the ICM conversion can be recovered is still unclear.  
However, the ICM representation allows us to have an appropriate circuit to {\em start} the optimisation procedure.  
Without the ICM form, the initial topological circuit would be undefined (as every $T$-gate requires a quantum correction 
50\% of the time, and one cannot predict which corrections will be necessary before computation begins).  Therefore  
circuit compression would be restricted to individual $T$-gates or be deliberately wasteful 
(by including space/time regions in the 
computation for corrections regardless of whether or not they are needed) or attempt to optimise as many of the 
$2^{\#T\mbox{-gates}}$ combinations of circuits as possible.  Neither of these options is desirable.\\

\section{Time-ordered ICM operations}

A second significant issue that needs to be considered when determining at what level quantum circuits are converted into 
ICM form and optimised is related to the classical information required in the circuit and when that information 
becomes available.  As described in the previous sections, various measurement results in the ICM representation dictate 
which basis subsequent measurements are performed in.  These basis measurements are always in either the $X$ or 
$Z$-basis and are therefore related through a Hadamard operation that can be implemented using code deformation \cite{F12}
rather than additional teleportations, without destroying the ICM structure.  However, when optimising a circuit globally, 
extraordinary care must be taken to ensure that classical information from these measurements is obtained as soon as 
possible and without disrupting the time-ordering required from the ICM representation.  

Rules for topological compression do not take into account the aspects of time ordering.  The example shown in 
Figure \ref{fig:topological} does not have time ordering (as the entire circuit is based on Clifford elements) and so 
no further constraints need to be imposed when performing optimisation.  However, for general circuits, this will not be 
the case and further constraints need to be placed on any optimisation technique.  

As we are still developing compression techniques and optimisation software for topological quantum circuits, it remains 
unclear how these constraints will effect the level to which circuits can be optimised.  Manual techniques were used 
in benchmarking Shor's algorithm \cite{DSMN13}, where circuits were optimised at the level of individual state distillation circuits 
and $T$-gates \cite{FD12,FDJ13}.  The ICM form allows a circuit or topological 
structure to be defined for an arbitrarily large circuit, but it may 
be inefficient or undesirable to compress and optimise extremely large circuits.  Instead it may be more preferable to 
optimise individual compound gates, or small subroutines that are routinely used.  Until we have a more concrete formalism 
for optimisation, particularly for the case of topological circuits, the level to which the ICM form is used for the high-level 
algorithm remains an open question.

\section{Conclusion}
In this paper we have introduced software to convert a high-level quantum algorithm into a fault-tolerant representations that 
consists of a set of qubit initialisations, a large CNOT array and measurements in the $X$- and $Z$-basis.  This allows 
us to design a circuit compatible with all fault-tolerant protocols for a large range of error correcting codes whose only 
non-deterministic element is the basis choices for measurement.  The ICM representation allows us to generate a canonical 
geometric description that fully describes the implementation of the algorithm using topological codes that forms the 
basis of all current hardware models.  The ICM representation for quantum circuits is directly related to algorithmically specific graph states that can be produced by mapping a quantum circuit to a standard 2-dimensional cluster state and performing 
all Clifford basis measurements in the algorithmic specification prior to computation.  Our work extends this to fully error corrected circuits, their 
necessary decomposition and ancillary protocols; it also can be used to connect work in higher level circuit synthesis and optimisation to the reality of implementing and optimising these algorithms on actual physical hardware models.

\section{Acknowledgements}
SJD acknowledges support from the JSPS Grant-in-aid for Challenging Exploratory Research. 
IP acknowledges support from BAYFOR grants BayIntAn\_Uni\_Passau\_2012\_21 and BayIntAn\_Uni\_Passau\_2014\_52.






\bibliographystyle{unsrt}

\section{Appendix}
\subsection{Gate Primitives and Decompositions:}
\begin{figure*}[ht!]
	\includegraphics[width=\textwidth]{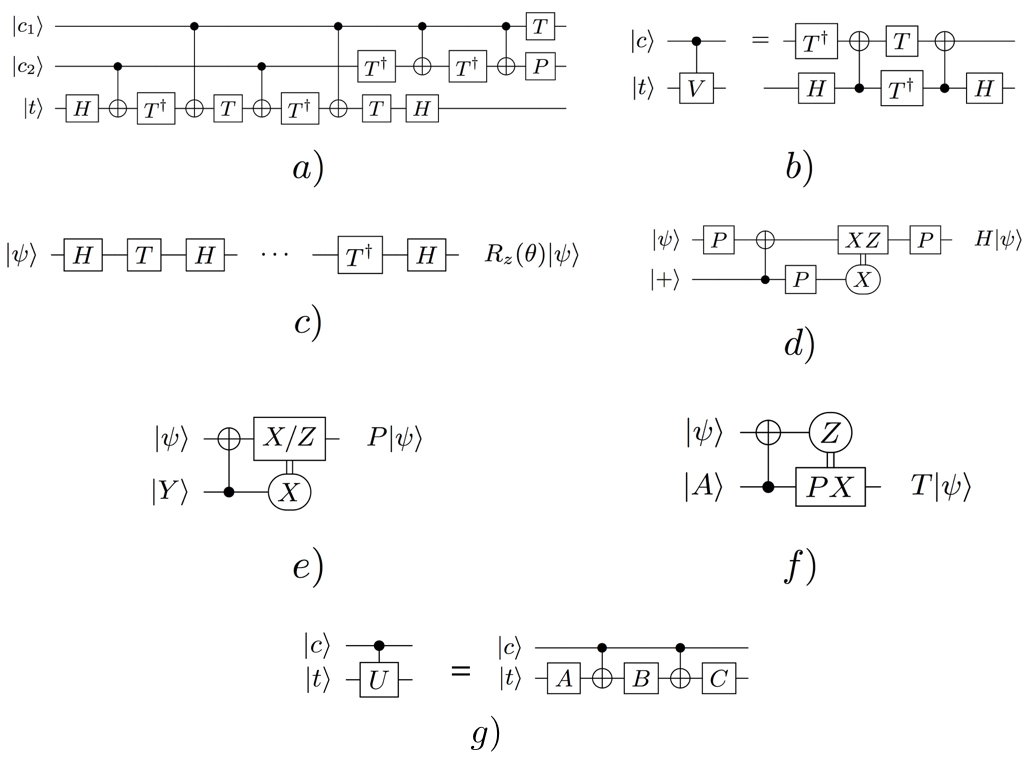}
\caption{{\bf Circuit decompositions. a)} Toffoli gate using CNOT, $T$, $T^\dagger$ and $H$ gates \cite{NC00}. {\bf b)} Controlled-$V$ gate.  {\bf c)} Arbitrary $R_z(\theta)$ gate decomposed in terms of $H$- and $T$-gates.  The length of the sequence, $L$ is related to the approximation error of the gate, $\epsilon$, $L = O(\log(1/\epsilon))$.  {\bf d)} Hadamard gate using $P$-gates.  The $X$-basis measurement determines a $Y=XZ$ Pauli correction.
{\bf e) and f)} Teleported rotational gates using the magic states $\ket{A}$ and $\ket{Y}$. These ancillary states are injected at high error, encoded 
and then purified using ancillary protocols \cite{BK05+}.  The correction for the $P$-gate is a Pauli $Y$-gate and can be tracked, while the 
correction for the $T$-gate requires a subsequent $P$-gate that must be applied to the actual quantum data. {\bf g)} is the decomposition of 
an arbitrary controlled unitary, where the single qubit gates, $A$, $B$ and $C$ are further decomposed into Clifford + $T$ gates. }
\label{circ:toffoli}
\end{figure*}

We begin with a set of circuit decompositions from higher level circuit primitives into an appropriate Clifford + $T$ 
set of gates, compatible with fault-tolerant error correction.  Our compiler uses the following high level circuit primitives: CNOT, Toffoli, 
Controlled-$V^{(\dagger)}$ ($V = \sqrt{X}$), Hadamard and other arbitrary single qubit gates.  
The Toffoli and Controlled-$V^{(\dagger)}$ gates 
are replaced with standard decompositions in terms of the Clifford + $T$ library as shown in Figure \ref{circ:toffoli}a) and Figure \ref{circ:toffoli}b).  For the Toffoli gate, 
we use the original decomposition containing seven $T$-gates \cite{NC00}, and the software will be periodically updated with other decompositions with reduced $T$-counts and $T$-depths \cite{J12,J13}.  

Hadamard gates are decomposed in terms of teleportation 
circuits using the ancilla state $\ket{Y}$, as shown in Figure \ref{circ:toffoli}d).  For topological implementations there are methods of 
applying the Hadamard using code deformation \cite{F12}, which allow for the ancilla state $\ket{Y}$ to be repeatedly used. Otherwise it needs to be distilled each time such a state is required \cite{J13+}. We choose not to utilise this technique because the code deformation protocol 
acts somewhat like a black box, and optimisation has to occur \emph{around} these objects.  While utilising a teleported 
Hadamard requires more resources (as $\ket{Y}$ states can no longer be recycled), this maintains the ICM form for the 
entire circuit.  Adding in black box elements (such as code-deformed Hadamards or other circuit substructures) may prove to 
be beneficial in future optimisation algorithms.  However, at this stage a global ICM representation is a good starting point.   

For arbitrary single qubit rotations, we utilise the algorithm of Ross and Selinger \cite{RS14}.  This algorithm comes close to achieving the theoretical lower 
bound for approximating arbitrary $Z$-rotations in terms of Clifford + $T$ gates up to a specified error, $\epsilon$ \cite{DN06}.  Recent 
techniques in probabilistic circuits have demonstrated a scaling for approximating $Z$-axis rotations \emph{better} than this 
lower bound, but they reintroduce probabilistic corrections that this work is attempting to eliminate \cite{BRS15,BRS15+}.  
Arbitrary axis rotations are then 
achieved via a standard Euler angle decomposition, with $X$-axis rotations approximated via $Z$-axis rotations and Hadamard gates, 
$R_{\vec{n}}(\omega) = R_z(\alpha)HR_z(\beta)HR_z(\gamma)$.  
\\
\\
\subsection{Selective Source and Destination Teleportation: Eliminating dynamic corrections}
In many fault-tolerant, error corrected models, $P$- and $T$-gates are achieved via the teleportation circuits shown in 
Figure \ref{circ:toffoli}e) and Figure \ref{circ:toffoli}f).
These gates are intrinsically probabilistic, and with 50\% probability may apply the 
gates $P^{\dagger}$ or $T^{\dagger}$ instead.  For the $P$-gate, this indeterminism is not a problem, as the $Z$ gate is used 
as a correction ($ZP^{\dagger} = P$); this can be tracked by updating the Pauli frame \cite{PDNP14}. However, 
the correction for the $T$-gate is a 
$P$-gate ($PT^{\dagger} = T$), which can not be classically tracked and therefore must be applied using active quantum 
circuitry.  As this correction occurs with a probability of 50\%, the initial quantum circuit (consisting of many $T$-gates when 
fully decomposed) cannot be assumed and optimised prior to computation.  

In a result from Fowler \cite{F12+}, the following trick can be used to construct deterministic circuitry regardless of this probabilistic teleportation and transfer the indeterminism of the global circuit to basis choices for qubit measurements after the deterministic circuit is executed.  This is similar to measurement based quantum computation \cite{RB01}, where a universal, 
algorithmically independent, resource state is constructed and measurement outcomes inform basis choices for subsequent 
measurements.  

\begin{figure*}[ht!]
	\includegraphics[width=\textwidth]{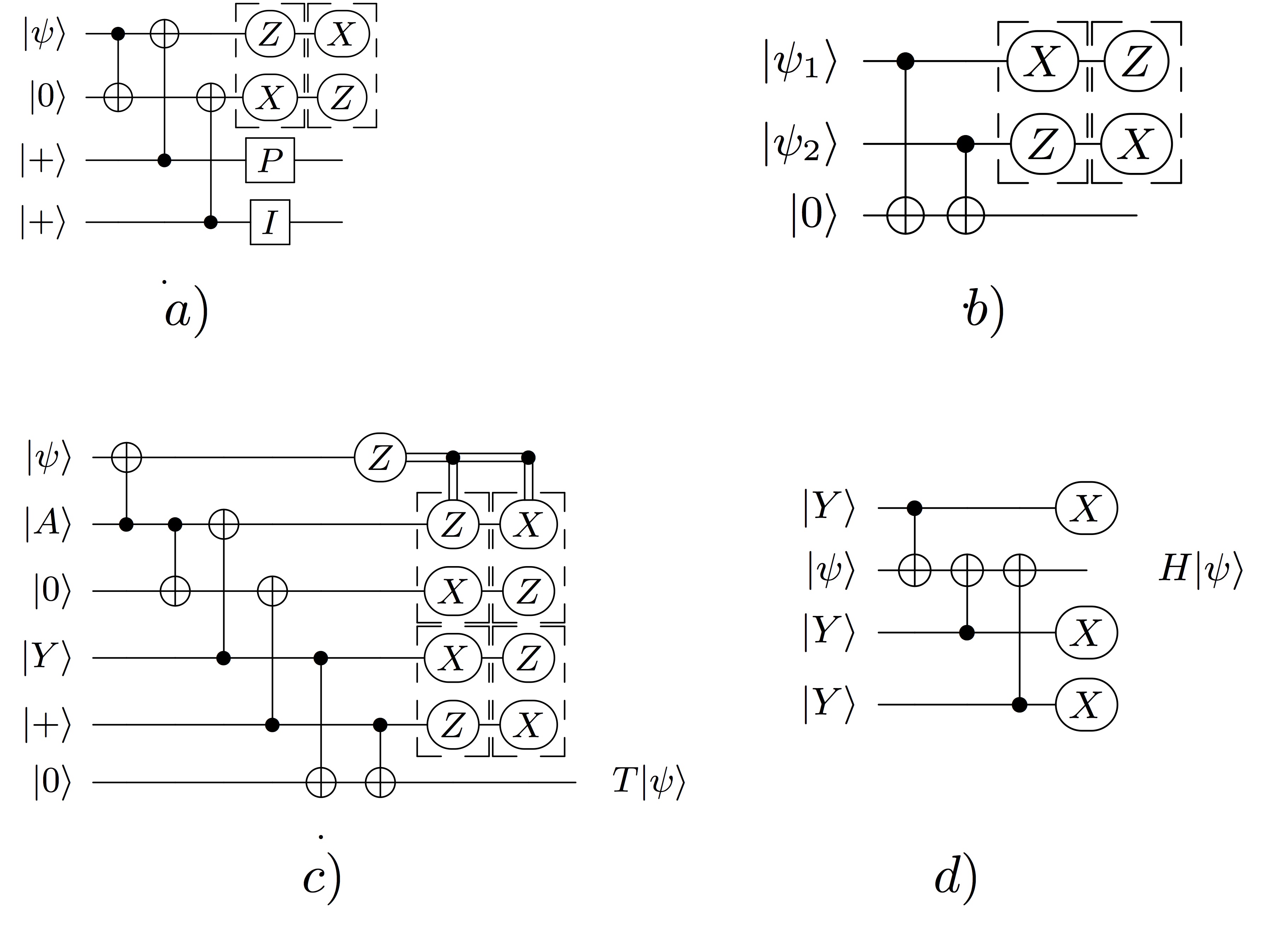}
\caption{{\bf Teleportations: a)} Selective destination; {\bf b)} Selective source \cite{F12+} and the combination of the two 
{\bf c)}, with the teleportation circuit [Figure \ref{circ:toffoli}d)] to produce 
deterministic circuitry for a $T$-gate. {\bf d)} 
Hadamard gate converted to ICM form.  Note that there are Pauli corrections based on all three $X$-measurements 
which are classically tracked.  Dotted boxes represent possible basis {\em choices} that chooses the target qubit (source qubits) for 
teleportation and in Figure d) is controlled based on the first $Z$-basis measurement.}
\label{circ:selective}
\end{figure*}

The two relevant circuits are known as selective source and selective destination teleportation, illustrated in Figure \ref{circ:selective}a) and Figure \ref{circ:selective}b).

The circuit in Figure \ref{circ:selective}a) allows us to selectively teleport a state to a qubit line where a $P$-gate is waiting to perform 
a correction or not, depending on the pattern ($Z/X$) or ($X/Z$) of the chosen measurements.  Once such a choice is made, the 
circuit in Figure \ref{circ:selective}b) allows us to selectively choose one of two source qubits to teleport to the output.  
These two circuits, combined, and commuting the $P$-gate correction through the control line in Figure \ref{circ:selective}a) allows 
us to perform a $T$-gate with deterministic circuitry [Figure \ref{circ:selective}c].  The original $Z$-measurement for the 
$T$-gate teleportation is now used to classically control a set of $X$- and $Z$-basis measurements on the other 
qubits (the dotted boxes represent the two basis \emph{choices}).  The $P$-gate correction can be commuted through the control of a CNOT gate and rotates one of the $\ket{+}$ qubits 
to the $\ket{Y} = P\ket{+}$ state.  

This basic element represents the building block of the ICM representation for an an arbitrary, high level, quantum circuit.  The structure is such that we have a vertical line of qubit initialisations ({\bf{I}}), followed by an array of only CNOT gates ({\bf C}) and then by a time-ordered series of either $X$- or $Z$-measurements ({\bf M}).

The Hadamard gate, using $\ket{Y}$ state resources, as shown in Figure \ref{circ:toffoli}b), can also be expressed in terms of teleportation circuits and hence be put into ICM form \cite{NC00}.
In Figure \ref{circ:selective}d), there are classical corrections based on the three $X$-basis measurements.  However, these corrections 
are all Pauli corrections and can be classically tracked. 
\\  

\subsection{Adding in Fault-tolerant Protocols:}
Based on the circuit decompositions shown so far, we take a high level quantum circuit, that may already be optimised with respect to a certain set of metrics. Additionally, we assume a universal gate set that is not automatically compatible with fault-tolerant error correction and we first convert it into a Clifford + $T$ representation, useful for both standard topological encoding and CSS concatenated encoding.  
Once this decomposition is done, we introduce selective source and destination circuits to place the entire circuit into an ICM form.  
This form of the higher level circuit is time optimal, given the details of the decomposition to the Clifford + $T$ level (i.e. the $T$-depth 
before introducing selective source and destination circuits determines the run-time of the computation, reducing $T$-depth at this 
level will reduce computation time \cite{F12+}).  We do not consider optimisation at this higher level, as we do not focus on determining the best structure of the high-level circuit in terms of reduced $T$-depths.

At the initialisation layer, we have a selection of four basis states, $\{\ket{0}, \ket{+}, \ket{Y}, \ket{A}\}$.  The $\ket{0}$ and 
$\ket{+}$ states can be initialised fault-tolerantly, intrinsically, with all relevant QEC codes for large scale hardware architectures.  
The states $\ket{A}$ and $\ket{Y}$, in general, cannot be.  This is particularly true for QEC models currently considered for large-scale 
hardware architectures, the topological surface and Raussendorf codes.  These states must be injected into the computer at 
high error and then recursively distilled.  For the $\ket{A}$ state, we utilise 15 low-fidelity states and then run a distillation circuit based 
on the [[15,1,3]] Reed-Muller code.  This, when successful, will reduce the error from $O(p)$ to $O(p^3)$; if this is still not sufficient, the process is concatenated, and 15 of these output states are used in a further distillation circuit.  For the $\ket{Y}$ state, a similar procedure is performed, this time utilising seven resource states and the [[7,1,3]] Steane code.   For states with input error rates of $p$, each 
of these circuits fail with probability $O(p)$ \cite{BK05+}.  

\begin{figure*}[ht!]
	\includegraphics[width=0.8\textwidth]{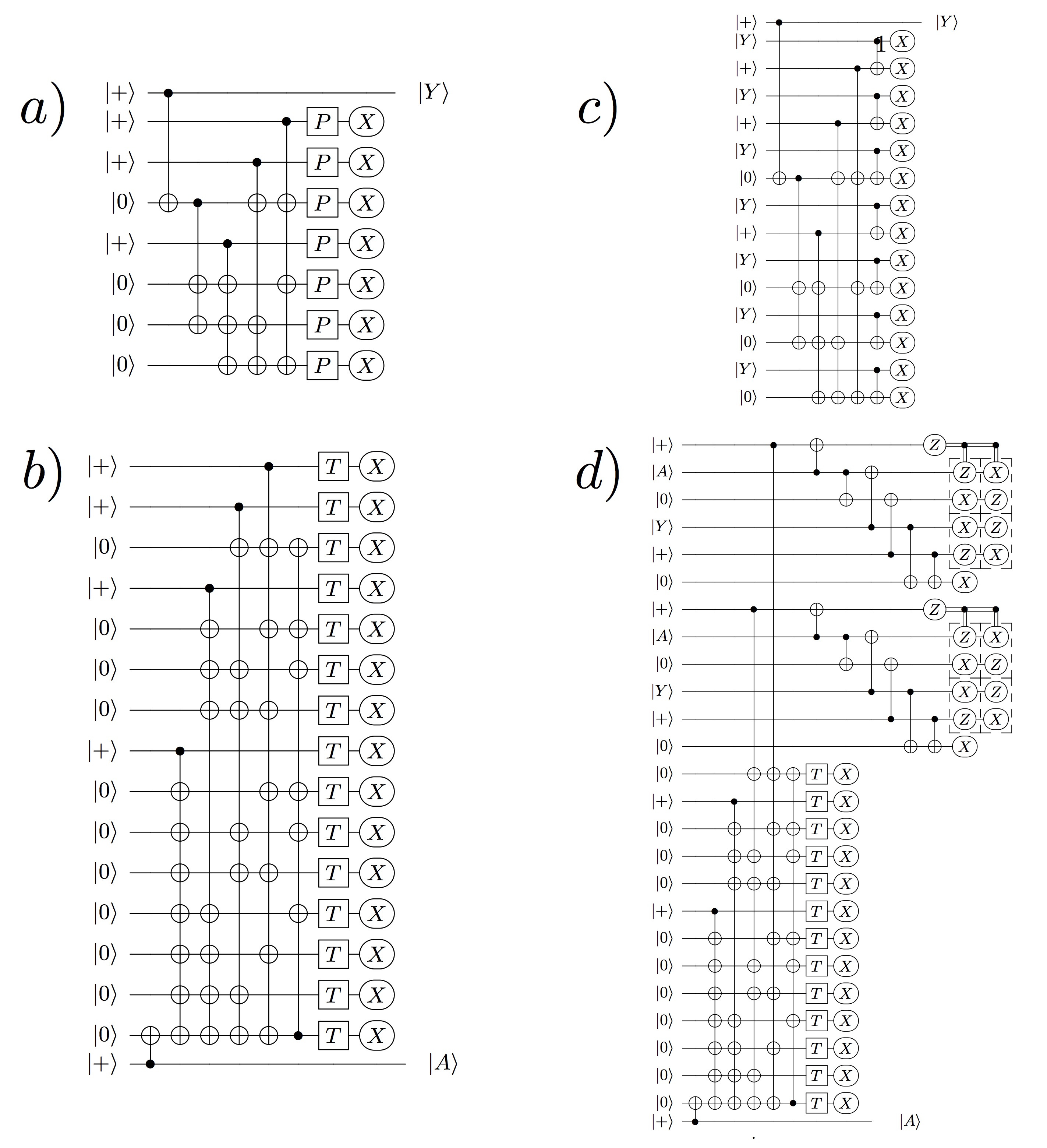}
\caption{{\bf Distillation circuits and their ICM representations.} State distillation circuits for the {\bf a)} $\ket{Y}$ and {\bf b)} $\ket{A}$ states.  When all $X$-measurements produce a trivial 
syndrome, the output decreases infidelity from $O(p)$ to $O(p^3)$, for an infidelity of $p$ for each of the $P$ gates. S
ICM representations for the {\bf c)} $\ket{Y}$ and {\bf d)} $\ket{A}$ states in time-optimal format.  For 
$\ket{A}$ state distillation we only illustrate the replacement for the first two $T$-gates to save space.  There is no 
time ordered sequence of measurement for $\ket{Y}$ state distillation, while the time sequencing for 
$\ket{A}$ state distillation is independent for each $T$-gate, indicative of a circuit with $T$-depth one.}
\label{circ:distillation}
\end{figure*}

Examples of these circuits are shown in Figure \ref{circ:distillation}a) and Figure \ref{circ:distillation}b), which are well known in the literature.  In Figures \ref{circ:distillation}c) and Figure \ref{circ:distillation}d) 
we convert both of them into their own ICM representations.  

\begin{figure}[ht!]
	\includegraphics[width=0.4\textwidth]{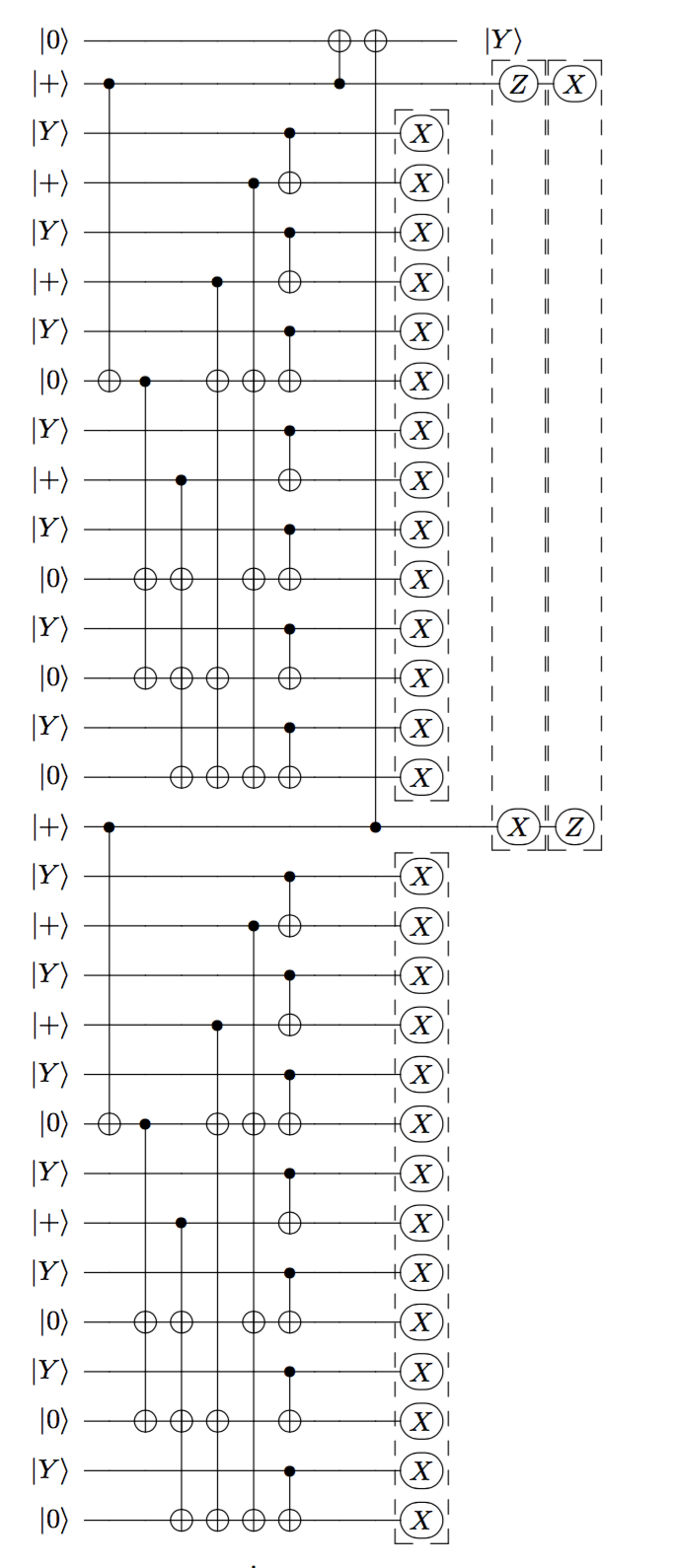}
\caption{Selective source teleportation is used to chose one of the distilled $\ket{Y}$ states in the event that a 
distillation circuit fails.  The last bank of $X$- or $Z$-basis measurements is determined by the $X$-measurements 
of each distillation circuit.  For injected states with error $p$, this circuit {\em will not} produce a valid output with 
probability $O(p^2)$, this can be decreased arbitrarily by including more distillation circuits and selective source 
teleportations.}
\label{fig:distillation3}
\end{figure}

These structures are in the ICM form, and can be used to replace any $\ket{A}$ or $\ket{Y}$ state in the initialisation 
stage for a higher level circuit.  A single level of state distillation is illustrated, being intended to take the error associated with 
the output state from an initial $O(p)$ to $O(p^3)$.  If this is not sufficient, then each $\ket{A}$ or $\ket{Y}$ state in these circuits 
are themselves replaced with identical structures.  Thus the ICM form is maintained and takes the error rate of each $O(p)$ injected state, 
to an $O(p^9)$ distilled output.  This recursion can continue as much as required (but often two levels are sufficient for large-scale algorithms) \cite{DSMN13,FMMC12}. State distillation only succeeds when 
no errors are detected by the final $X$-basis measurements.  Hence, for states injected with an error rate $p$, the circuit 
will fail with probability $O(p)$.  To maintain a deterministic array of CNOT gates in the ICM representation, we use the same 
trick with selective source and destination teleportation and duplicate the distillation circuits.  This is illustrated in Figure 
\ref{fig:distillation3}. 

To protect against failures of a distillation circuit, we utilise selective source teleportation on multiple copies.  The $X$- and $Z$-basis measurements on the teleportation circuits are predicated on the $X$-measurements returning a trivial 
syndrome on at least one of the distillation circuits.  As a consequence, the probability of {\em not} producing a distilled 
$\ket{Y}$ state  is reduced from $O(p)$ to $O(p^2)$.  This procedure can be repeated to reduce the probability of failure 
arbitrarily.  To reduce resources, if multiple distillation circuits succeed, outputs can be routed using selective source and 
destination teleportations to other regions of the computer where distillation has failed.

The initial decomposed ICM form, combined with these distillation structures for $\ket{A}$ and $\ket{Y}$ states now allows to 
represent an arbitrary high level algorithm into an ICM representation, incorporating all necessary fault-tolerant ancillary protocols.   

\section{Compiler Code}
The compiler source code is released under the Microsoft Reference Source License (Ms-RSL, \url{http://referencesource.microsoft.com/license.html}) at \url{https://github.com/alexandrupaler/icmconvert}. A first application of the compiler exists at \url{https://github.com/alexandrupaler/tqec}, which is gradually updated with respect to ICM compilation \cite{PDF16}, too.

The ICM compiler consists of three components whose interaction will be explained and exemplified in the following. Converting an arbitrary circuit into ICM requires the decomposition of the non-ICM quantum gates, and there are two possibilities: either a gate can be exactly decomposed into ICM (first component: the \texttt{convertft} tool), or it needs to be approximated first into gates that are easily ICM representable (second component: the \texttt{decompose} and \texttt{processraw} tools). Irrespective of the chosen method, the decompositions are stored into and retrieved from a text database accessed using the methods of the \texttt{databasereader} class (third component).

The database contains three types of decompositions. The \texttt{icmdist} decomposition refers to subcircuits used to replace state injections in an ICM circuit. The \texttt{nicm} decomposition, used for example for Toffoli gates, represents the decomposition of a (non-)ICM gate into non-ICM elements (gates, measurements). Finally, \texttt{icm} stands for ICM decompositions of non-ICM gates.

From an arbitrary quantum circuit towards an ICM one, the database plays a central role. In general, each quantum gate is expressed as a unitary matrix and has a name, such that a quantum circuit is a list of quantum gate names and the qubits operated on. The nicm decompositions are obtained by the \texttt{decompose} tool, and undecomposed quantum gates are specified in a text file, where the complex matrix entries are written in polar form. For example, the Hadamard gate (although a direct ICM representation is known) is
{\scriptsize
\begin{verbatim}
1
Hadamard
0.70710678118 0
0.70710678118 0
0.70710678118 0
- 0.70710678118 0
\end{verbatim}
}

The first line of the file indicates the number of specified gates (e.g. $1$). Each gate specification consists of name (e.g. Hadamard) and four lines for the complex numbers of the unitary matrix (e.g. radius $0.70710678118$ and angle $0$). The \texttt{decompose} tool reads such a file and outputs corresponding nicm database entries.

After decomposing single qubit unitaries and storing their nicm decompositions into the database, the \texttt{processraw} tool takes the input circuit and replaces all occurences of nicm gates (e.g. Toffoli) with their corresponding database entries. Thus, if in the input circuit appears the name \texttt{toffoli}, its occurence is replaced by the decomposition stored into the database. For example, the nicm Toffoli gate decomposition is 

{\scriptsize
\begin{verbatim}
=toffoli
nicm
0
WIRE WIRE WIRE CTRL WIRE WIRE WIRE CTRL WIRE CTRL WIRE CTRL TGATE
WIRE CTRL WIRE WIRE WIRE CTRL WIRE WIRE TGATE TGT TGATE TGT PGATE
HGATE TGT TGATE TGT TGATE TGT TGATE TGT TGATE HGATE WIRE WIRE WIRE
\end{verbatim}
}

The name of the decomposed gate is specified on the first line after the '=' character. The second line (e.g. nicm) indicates the decomposition type, and the third line the number of required ancillae (e.g. zero). The convention is that for a controlled gate, as the Toffoli, the first qubits are the controls and the last ones are the targets. The final three lines represent the Toffoli decomposition. For each database entry the decomposition elements (e.g. WIRE, TGATE) are encoded as strings defined in \texttt{gatenumbers.h}.

The intermediate circuit obtained after executing \texttt{processraw} is finally processed by \texttt{convertft}. The tool outputs the ICM circuit description (\texttt{.circ} file), a geometrical description (\texttt{.geom} file) and a Postscript representation of the ICM circuit (\texttt{.ps} file). Furthermore, \texttt{convertft} takes as a parameter the number of distillation rounds to be included into the ICM circuit. The functionality of this tool is illustrated using the icm decompostion of the $T$ gate.
{\scriptsize
\begin{verbatim}
=TGATE
icm
1
EMPTY AA
c 2 1
MZ EMPTY
\end{verbatim}
}

The $T$ gate requires a single ancilla, and because of its ICM form the decomposition consists of the same three regions. The fourth line in the listing indicates the initialisations of the qubits: the initialisation of the first qubit is left unchanged (EMPTY) and the second qubit is initialised into $\ket{A}$ (AA). The lines starting with the character \texttt{c} represent CNOTs (e.g. \texttt{c 2 1} is a CNOT controlled by the second qubit and targeting the first qubit). The last decomposition line specifies measurement types: the first qubit is measured in $Z$ (MZ), and the second qubit is left unmeasured (EMPTY).

After replacing in the intermediate circuit each occurence of TGATE with the icm decomposition, \texttt{convertft} can include distillation sub circuits into the ICM output. For example, the $\ket{A}$ state distillation circuit is specified as ($\backslash$ indicates a continuation of the same text line)
{\scriptsize
\begin{verbatim}
=AA
icmdist
15
PLUS PLUS ZERO PLUS ZERO ZERO ZERO PLUS \
   ZERO ZERO ZERO ZERO ZERO ZERO ZERO PLUS EMPTY
c 16 15
c 1 3 5 7 9 11 13 15
c 2 3 6 7 10 11 14 15
c 4 5 6 7 12 13 14 15
c 8 9 10 11 12 13 14 15
c 15 3 5 6 9 10 12
MA MA MA MA MA MA MA MA MA MA MA MA MA MA MA EMPTY
\end{verbatim}
}

Distillations are included by replacing each occurence of the AA initialisation with the above specified circuit. The measurements in the $\ket{A}$ basis (MA) are not corresponding to the ICM definition offered in the introduction in the main text, and in a second step each MA occurence is replaced by a $T$ gate applied before an $X$ basis measurement. Finally, the $T$ gates are ICM decomposed.
{\scriptsize
\begin{verbatim}
=MA
nicm
0
TGATE MX
\end{verbatim}
}

The above $T$ gate decompositions did not include the selective source and destination teleportations, but these can be easily included by augmenting the TGATE database entry. In this case, the variable ancillae measurement types (effect of probabilistic corrections) are specified using the operations MXZ and MZX also defined in \texttt{gatenumbers.h}.
{\scriptsize
\begin{verbatim}
=TGATE
icm
5
EMPTY AA ZERO YY PLUS ZERO
c 2 1
c 2 3
c 4 2
c 5 3
c 4 6
c 5 6
MZ MZX MXZ MXZ MZX EMPTY
\end{verbatim}
}

The circuit description presented in the results section of the main text is a reformulation of the database entry format. The initialisation and measurement lines in the circuit description are compressed to a single line in the database entries, and instead of the \texttt{cnot} command, the database uses the \texttt{c} command. This fact shows that, once a circuit is ICM transformed, it can be easily stored into the database for future transformations to use it as a sub circuit (e.g. modular adders used in modular multiplications).

\end{document}